\documentclass[twocolumn]{emulateapj}
\usepackage{apjfonts}

\usepackage{natbib}
\usepackage{epsfig}
\usepackage{rotating}
\usepackage{hyperref}

\newcommand{\be}{\begin{equation}}
\newcommand{\ee}{\end{equation}}

\newcommand{\eg}{\emph{e.g.}}
\newcommand{\kms}{\mbox{km\,\ensuremath{\rm{s}^{-1}}}}
\newcommand{\ebv}{$E_{B-V}$}
\newcommand{\dib}{$\lambda$}
\newcommand{\cp}{C$_{60}^+$}
\newcommand{\bd}{BD+63\,1964}
\newcommand{\espadons}{ESPaDOnS}

\submitted{ApJL, accepted for publication, June 2017}

\shortauthors{Cordiner et al.}

\begin{document}

\title{Searching for interstellar C$_{60}^+$ using a new method for high signal-to-noise\\ HST/STIS spectroscopy}

\author{M. A. Cordiner\altaffilmark{1,2}, N. L. J. Cox\altaffilmark{3,4}, R. Lallement\altaffilmark{5}, F. Najarro\altaffilmark{6}, J. Cami\altaffilmark{7,8}, T. R. Gull\altaffilmark{1}, B. H. Foing\altaffilmark{9}, H. Linnartz\altaffilmark{10}, D. J. Lindler\altaffilmark{1,11}, C. R. Proffitt\altaffilmark{12}, P. J. Sarre\altaffilmark{13}, S. B. Charnley\altaffilmark{1}}


\altaffiltext{1}{NASA Goddard Space Flight Center, 8800 Greenbelt Road, Greenbelt, MD 20771, USA}
\email{martin.cordiner@nasa.gov}
\altaffiltext{2}{Department of Physics, Catholic University of America, Washington, DC 20064, USA}
\altaffiltext{3}{Anton Pannekoek Institute for Astronomy, University of Amsterdam, NL-1090 GE Amsterdam, The Netherlands}
\altaffiltext{4}{Universit\'e de Toulouse, UPS-OMP, IRAP, 31028, Toulouse, France}
\altaffiltext{5}{GEPI, UMR8111, Observatoire de Paris, 5 Place Jules Janssen, 92195, Meudon, France}
\altaffiltext{6}{Departamento de  Astrof\'{\i}sica, Centro de Astrobiolog\'{\i}a (CSIC/INTA), ctra. de Ajalvir km. 4, 28850 Torrej\'on de Ardoz, Madrid, Spain}
\altaffiltext{7}{Department of Physics and Astronomy and Centre for Planetary Science and Exploration (CPSX), The University of Western Ontario, London, ON N6A 3K7, Canada}
\altaffiltext{8}{SETI Institute, 189 Bernardo Ave, Suite 100, Mountain View, CA 94043, USA}
\altaffiltext{9}{ESA ESTEC SCI-S, Noordwijk, The Netherlands}
\altaffiltext{10}{Sackler Laboratory for Astrophysics, Leiden Observatory, Leiden University, PO Box 9513, NL 2300 RA Leiden, Netherlands}
\altaffiltext{11}{Sigma Space Corporation, 4600 Forbes Blvd., Lanham, MD 20706, USA}
\altaffiltext{12}{Space Telescope Science Institute, 3700 San Martin Drive, Baltimore, MD 21218, USA}
\altaffiltext{13}{School of Chemistry, The University of Nottingham, University Park, Nottingham, NG7 2RD, UK}

\begin{abstract}

Due to recent advances in laboratory spectroscopy, the first optical detection of a very large molecule has been claimed in the diffuse interstellar medium (ISM): \cp\ (ionized Buckminsterfullerene). Confirming the presence of this molecule would have significant implications regarding the carbon budget and chemical complexity of the ISM. Here we present results from a new method for ultra-high signal-to-noise (S/N) spectroscopy of background stars in the near infrared (at wavelengths 0.9-1~$\mu$m), using the Hubble Space Telescope Imaging Spectrograph (STIS) in a previously untested `STIS scan' mode. The use of HST provides the crucial benefit of eliminating the need for error-prone telluric correction methods in the part of the spectrum where the \cp\ bands lie, and terrestrial water vapor contamination is severe. Our STIS spectrum of the heavily-reddened B0 supergiant star \bd\ reaches an unprecedented S/N for this instrument ($\sim600$-800), allowing the detection of the diffuse interstellar band (DIB) at 9577~\AA\ attributed to \cp, as well as new DIBs in the near-IR. Unfortunately, the presence of overlapping stellar lines, and the unexpected weakness of the \cp\ bands in this sightline, prevents conclusive detection of the weaker \cp\ bands. A probable correlation between the 9577~\AA\ DIB strength and interstellar radiation field is identified, which suggests that more strongly-irradiated interstellar sightlines will provide the optimal targets for future \cp\ searches.

\end{abstract}

\keywords{ISM: molecules --- instrumentation: spectrographs --- Techniques: spectroscopic --- line: identification}

\section{Introduction}

The diffuse interstellar band (DIB) problem is the longest-standing puzzle in interstellar chemistry, seemingly impenetrable despite the dedicated efforts of astronomers and laboratory chemists since the early 20th century \citep{her95,sar06,sno14}. The DIBs manifest as broad spectroscopic absorption features in the optical to near-infrared (NIR) spectra of stars as their light passes through the diffuse interstellar medium, indicating the presence of a large quantity of (mostly carbonaceous), { unidentified} molecular material \citep[\eg][]{cor11}. \citet{foi94} discovered two DIBs at 9577~\AA\ and 9632~\AA\ that they assigned to \cp\ (ionized Buckminsterfullerene) based on a similarity with the absorption wavelengths seen in neon matrix spectroscopy \citep{dhe92,ful93}. Recently obtained gas-phase spectra of \cp$-$He$_n$ complexes (for $n\leq4$; \citealt{cam15,cam16,khu16}), show that the match between the interstellar and laboratory wavelengths is accurate to within the observational uncertainties (a fraction of an angstr{\"o}m). { The bare \cp\ absorption wavelengths were confirmed in an independent laboratory study by Spieler et al. (2017, submitted) using \cp\ cations embedded in He-droplets.}

If confirmed, the detection of interstellar C$_{60}^+$ will constitute a major breakthrough in interstellar chemistry and may provide, for the first time, an insight into the true scale of chemical complexity in the diffuse ISM. Measurement of the NIR electronic transitions of C$_{60}^+$ (in absorption) would also provide a unique complement to the discovery of mid-infrared C$_{60}$ and C$_{60}^+$ emission bands in circumstellar and interstellar environments \citep{cam10,sel10,ber13}.

Despite dedicated observational studies \citep{wal15,wal16,gal17}, the case for interstellar C$_{60}^+$ has not yet been proven beyond reasonable doubt. Based on the laboratory measurements, five absorption features are expected (at 9348.4, 9365.2, 9427.8, 9577.0 and 9632.1~\AA, with strength ratios 0.07:0.2:0.3:1.0:0.8; \citealt{cam16b}). \citet{gal17} were unable to confirm the presence of the weakest three features in a sample of 19 heavily-reddened Galactic sightlines observed from the ground. Instead of a constant ratio for the two strongest C$_{60}^+$ DIBs (\dib9577 and \dib9632), as expected for electronic transitions arising from a $^2A_{1u}$ ground vibronic state, \citet{gal17} found that the interstellar band ratio was highly variable among different lines of sight.  We note, however, that if the observed transitions involve lower levels above the ground state (such as the split levels arising from Jahn-Teller distortion), a variable ratio could potentially occur. Although \citet{wal16} claimed interstellar detections of all five \cp\ bands, the three weaker \cp\ bands are problematic as they fall in a wavelength region heavily obscured by absorption due to water vapor in the Earth's atmosphere \citep[see][]{gal00,gal17}.  Telluric correction methods for such weak interstellar absorption features are error-prone as a result of incomplete telluric line cancellation, and from the possible presence of weak (unseen) stellar lines in the telluric standard spectrum. To rigorously confirm the identification of interstellar C$_{60}^+$, high signal-to-noise observations of all five absorption bands are required, preferably from outside Earth's atmosphere.

In this article, we present the first ever ultra-high S/N, high resolution stellar/interstellar spectra obtained using the Hubble Space Telescope (HST) imaging spectrograph, targeting four of the five \cp\ features (\dib9349, \dib9365, \dib9428 and \dib9577) along a heavily reddened line of sight.  Very high continuum signal-to-noise ratios (S/N$>500$ per spectral channel) are commonly achieved using ground-based telescopes, but such sensitivity has not previously been obtained using (direct) HST spectroscopy due to the severe CCD fringing at red and near-IR wavelengths. After successfully employing a previously untested `STIS scan' observing mode to trail the target star along the spectrograph slit (crossing hundreds of CCD rows to facilitate fringe cancellation), unprecedented signal-to-noise ratios $\sim600$-800 have been obtained. These spectra have permitted the first search for weak interstellar absorption features in the NIR, unhindered by telluric absorption.

\section{Observations and data reduction}\label{obs}

To reliably detect the weaker \cp\ DIBs, a high-S/N interstellar spectrum is required, with minimal contamination from interloping stellar photospheric lines. The heavily-reddened (\ebv~=~1.01) B0\,I star \bd\ was selected due to its extremely strong DIBs --- the majority of DIBs are among the strongest for stars of comparable extinction \citep{tua00} --- as well as its early spectral type and relatively clean predicted spectrum in the vicinity of the \dib9365 and \dib9428 \cp\ bands. The lightly-reddened B0\,I star 69~Cygni (\ebv~=~0.14) was selected for comparison, to assist in the discrimination of stellar and interstellar features.

\subsection{HST STIS}
\label{stis}

Data were acquired over a single orbit for 69~Cyg and two orbits for \bd, using the G750M grating with a central wavelength of 9336~\AA\ (covering the range 9050-9610~\AA). { This covers four of the five known \cp\ bands; a change of grating tilt would be required to obtain the fifth band}.  We used the $52''\times0.1''$ slit, and the plate scale was $0.05''$ per pixel. Following peakup and focusing maneuvers, a series of STIS-scan exposures was performed for each star, trailing the star along the slit at a constant rate with the shutter held open. For 69~Cyg, six exposures were obtained: two long scans spanning up to 1000 detector rows and four shorter scans spanning $\sim300$ rows. The scan direction was alternated for successive exposures.  For \bd\ (observed in a separate visit after the optimal observing sequence had been established), eight identical forward scans were obtained spanning 700 detector rows each, towards the upper part of the CCD. A pair of flat-field exposures (contemporaneous fringe flats) was obtained in sequence following each pair of target star exposures, and 5-6 additional pairs of flats were obtained to fill in the remaining orbital time during occultation. Exposures of a Pt/Cr-Ne lamp were obtained during each orbit for wavelength calibration. Subsequent conversion from vacuum to air wavelengths was performed using the \citet{mor00} formula. 

The STIS CCD suffers from charge transfer inefficiency (CTI) during readout, as a result of long-term radiation damage. The raw frames were corrected for CTI using the procedure outlined by \citet{and10}\footnote{see http://www.stsci.edu/hst/stis/software/analyzing/scripts/pixel\_based\_CTI}, which significantly reduces the severity of bad CCD columns. Standard dark current, bias, geometric distortion and wavelength corrections were performed using the {\sc Iraf} {\tt calstis} package \citep{hod98}. For \bd, the four trailed science exposures in each orbit were sufficiently well matched to allow statistical cosmic ray rejection using {\tt ocrreject}. This was not possible for 69~Cyg due to imperfect registration of the exposed rows between each frame, which may have occurred due to errors in the guide star tracking, as well as possible exposure timing errors.

\begin{figure}
\includegraphics[width=\columnwidth]{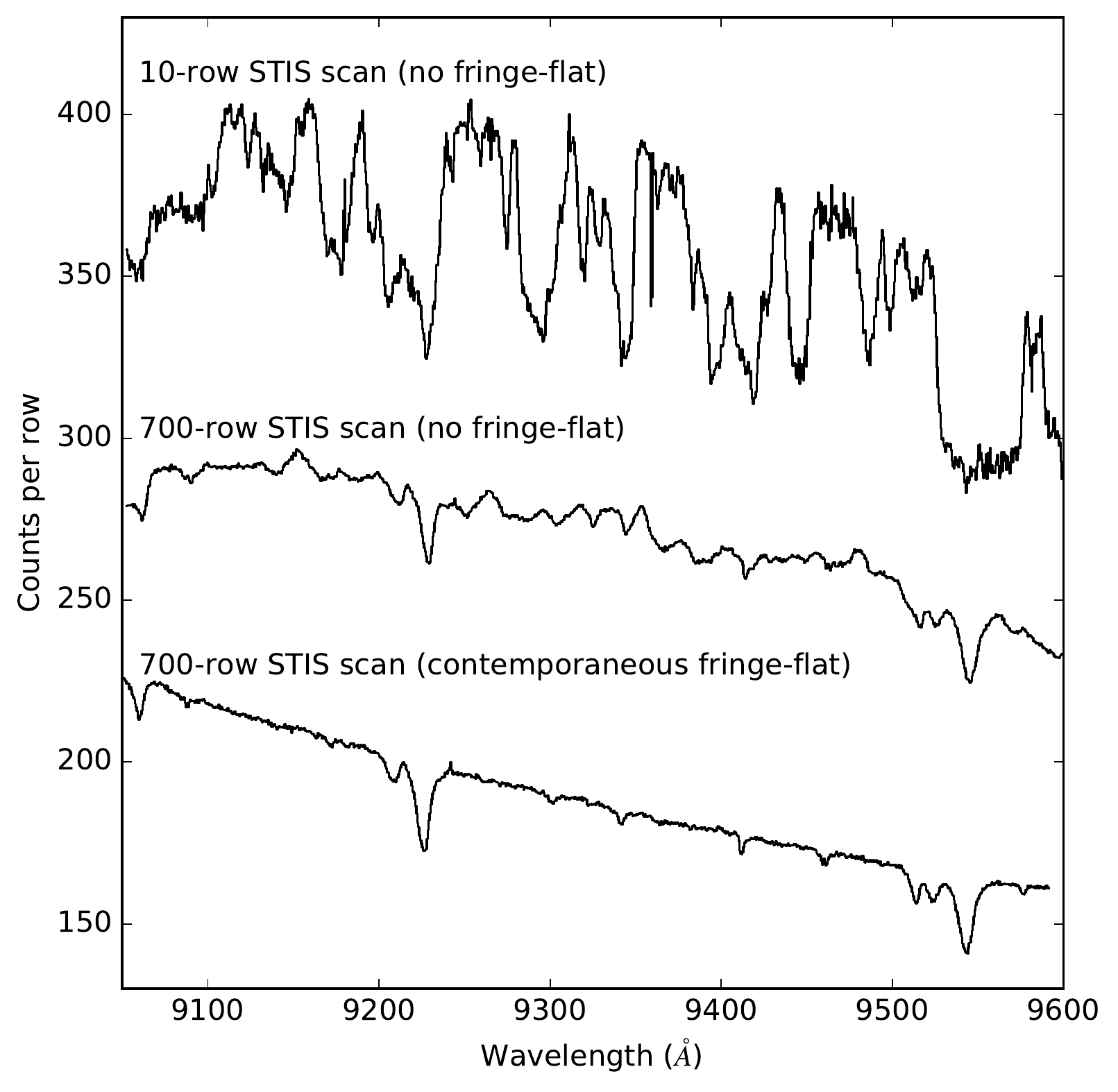}
\caption{HST spectra of \bd\ using three different scanning/flat-fielding schemes (with additive vertical offsets). Top: standard STIS spectroscopic acquisition and reduction (extracted over 10 dispersion rows), showing severe CCD fringing. Middle: a substantial reduction in fringe amplitude is achieved by STIS scanning (extracted over 700 dispersion rows). Bottom: the combined result of STIS scanning and flat fielding using a contemporaneous (in-orbit) fringe flat.\label{fig:fringe}}
\end{figure}

\begin{figure*}
\includegraphics[width=0.33\textwidth]{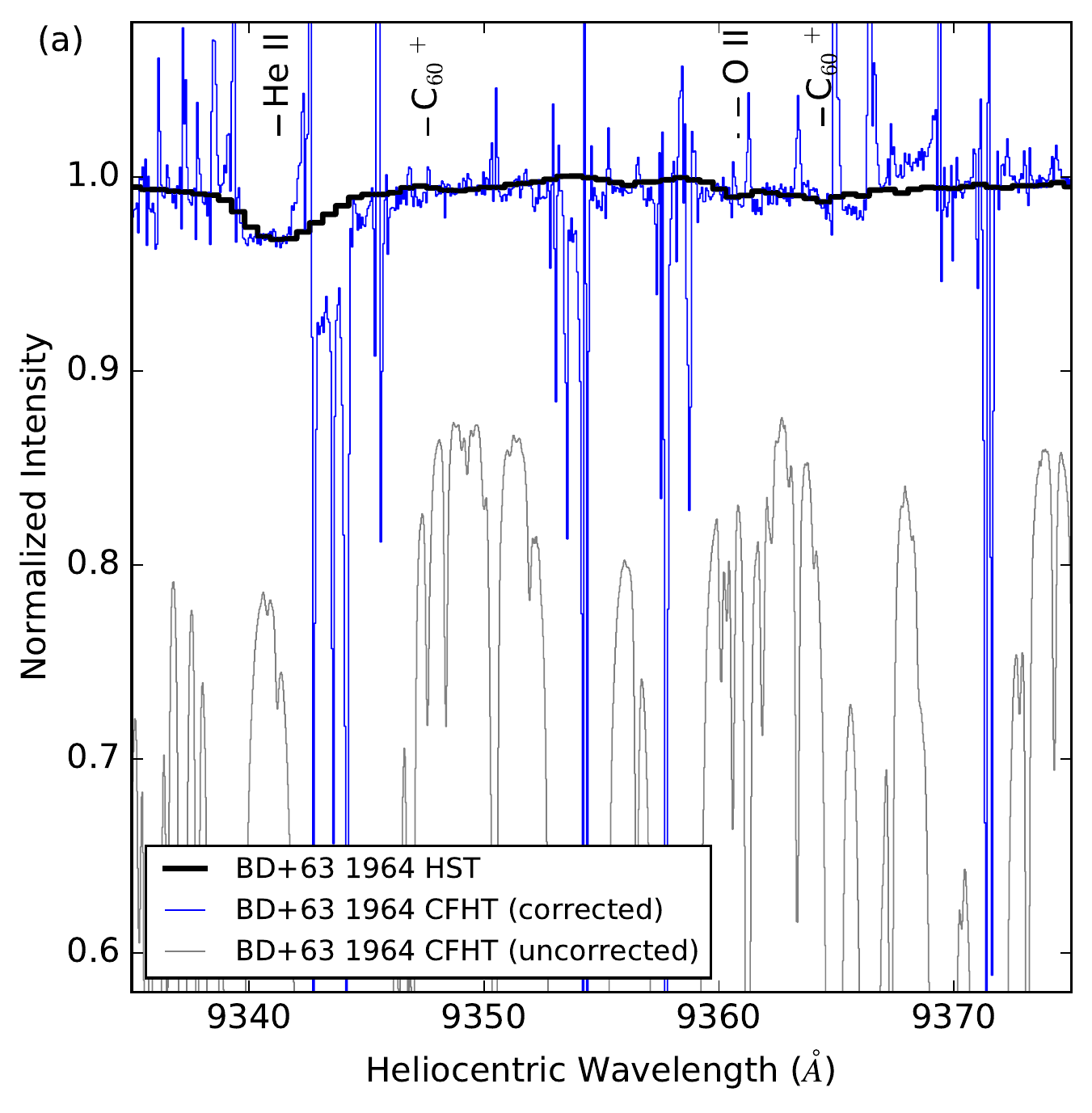}
\includegraphics[width=0.33\textwidth]{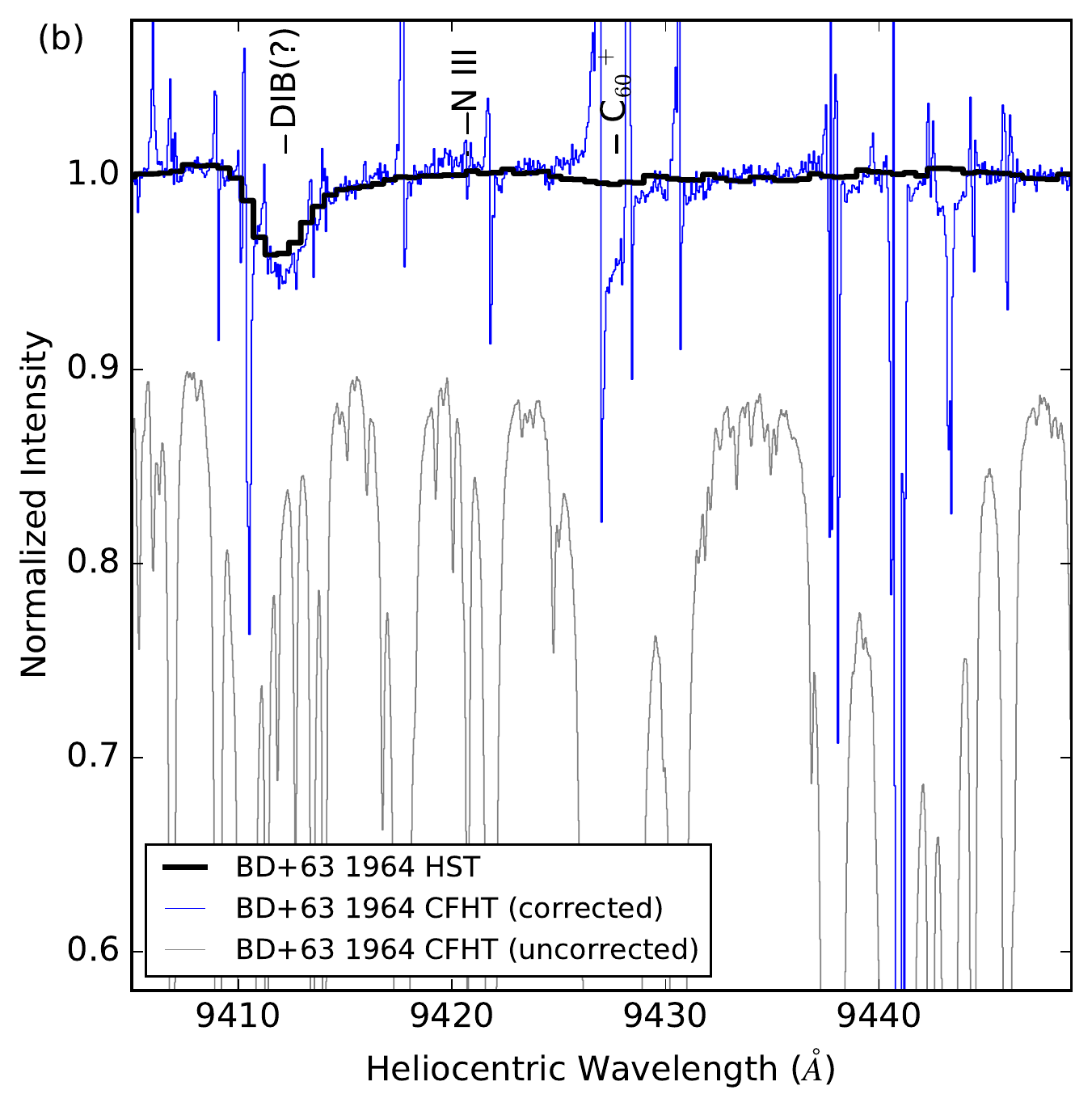}
\includegraphics[width=0.33\textwidth]{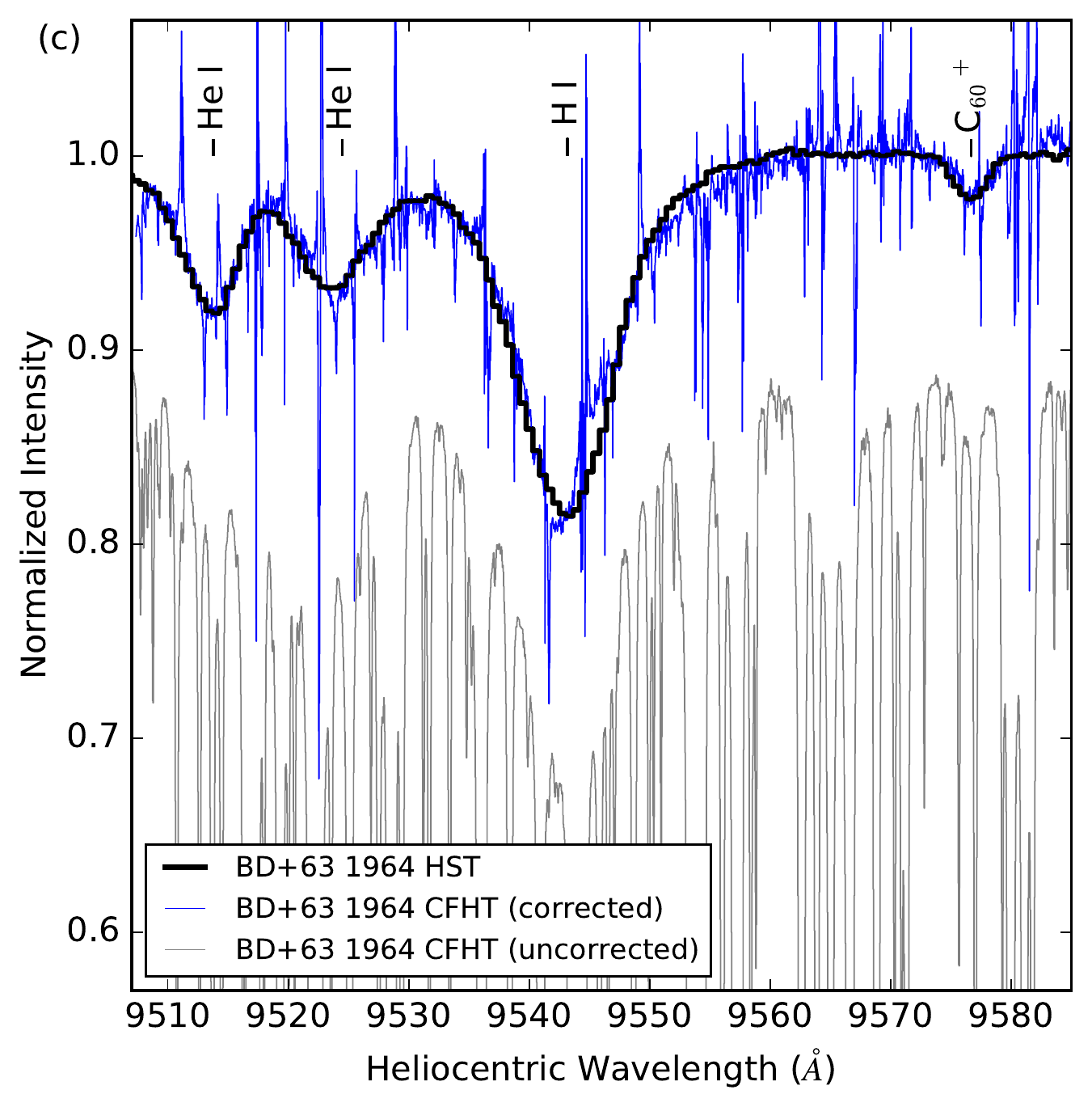}
\caption{Comparison of HST/STIS and CFHT/\espadons\ spectra of \bd, normalized with a linear continuum. { CFHT spectra were telluric-corrected using a TAPAS model; the uncorrected CFHT spectra (dominated by telluric absorption) are shown in gray with a vertical offset of $-0.1$.} Artifacts in the corrected CFHT spectra are primarily due to incomplete cancellation around the strongest telluric lines. The match between the strengths of the H and He lines in panel (c) demonstrates { close consistency of the STIS and \espadons\ calibrations}.   \label{fig:hstcfht}}
\end{figure*}

Spectral extraction was performed by summing the counts along CCD columns. Bad pixels and residual cosmic rays were rejected based on their standard deviation (using a $3\sigma$ threshold), after first normalizing the individual CCD rows by their median levels. Such row normalization was required due to variations in the illumination function along the slit, resulting from slight variations in the slit width and telescope scan rate.  Scattered light subtraction was performed with a low-order fit to the light under the two occulting bars (positioned 1/3 and 2/3 way along the slit).  Extracted Pt/Cr-Ne lamp line widths were consistent with the nominal instrumental resolving power of 10,000 (30~\kms).

The STIS CCD suffers from severe fringing due to internal reflections at wavelengths greater than 700~nm, where the chip starts to become transparent to incoming light. Fringe amplitudes are variable with wavelength, with a maximum of $\pm20\%$ near 900~nm (see \eg\ Figure \ref{fig:fringe}). The aim of our fringe cancellation strategy was to match as closely as possible the illumination pattern of the science exposure to that of the flat field, by STIS-scanning. This approach has the major advantages of (1) reducing the impact of individual bad pixels and cosmic ray hits, (2) allowing for orders of magnitude increases in the number of counts per science exposure, (3) eliminating the need to change the slit mechanism during observations, resulting in a consistent optical setup for the science and flat-field exposures.  

As shown in Figure \ref{fig:fringe}, scanning the exposed star across hundreds of CCD rows results in significantly reduced fringe amplitudes (a five-fold reduction in the spectral RMS in this case). Following correction using a contemporaneous fringe flat, we obtained apparently complete fringe cancellation down to the level of the statistical noise.

\subsection{CFHT \espadons}
\label{espadons}

For comparison with our STIS data, four 1000~s exposures of \bd\ were obtained over 2 nights in 2016 December, using the \espadons\ {\'e}chelle spectrograph of the Canada-France-Hawaii Telescope (CFHT). Observing conditions were good and the airmass was in the range of 1.41-1.49. The data were reduced by the automated Upena pipeline, which uses the Libre-ESpRIT data reduction software \citep{don97}. The reduced spectra cover the range 370-1048~nm at a resolving power of $\sim$80,000, and have a continuum signal-to-noise ratio $\sim500$ at 900~nm.

Due to the forest of atmospheric H$_2$O absorption lines in our STIS range, the \espadons\ spectra need to be telluric-corrected. The conventional approach of division by a standard star is unreliable for the measurement of weak DIBs due to the likely presence of stellar features in the standard star spectrum, which would introduce artifacts upon division.  We therefore adopted a transmission modeling approach, using synthetic atmospheric transmittances provided by the TAPAS website\footnote{http://ether.ipsl.jussieu.fr/tapas/} \citep{ber14}, adapted for Mauna Kea. A preliminary telluric correction was performed using the `rope length' method for weak to moderately strong lines \citep{rai12}, excluding from the length the spectral intervals corresponding to the central parts of the strongest lines. The preliminary correction was then interpolated in the most heavily contaminated intervals to remove residual artifacts. Next, the observed data were fitted to the convolved product of this spectrum and the TAPAS synthetic transmittance, allowing for variable airmass, refined wavelength correction and variable line-spread function. The observed data  were divided by this new, adjusted transmission spectrum to provide a new corrected spectrum. The second step was then iterated to convergence to produce the final corrected spectrum. { Despite our best efforts, telluric residuals still remain at the locations of the most heavily saturated lines, which may be due to time-variability of the telluric spectrum, scattered light residuals or small asymmetries in the \espadons\ instrumental response function.}

\subsection{Comparison between STIS and \espadons\ spectra}

The main purpose of these \espadons\ spectra was to verify the reliability of our nonstandard STIS observation and reduction procedures. A comparison between the HST/STIS and telluric-corrected CFHT/\espadons\ spectra is shown in Figure \ref{fig:hstcfht}. { Telluric-correction artifacts in panels (a) and (b) are severe, preventing a reliable comparison in the vicinity of the weaker \cp\ bands. However, in the less problematic range 9500-9600~\AA\ (panel c), the agreement between the two instruments is excellent, which} demonstrates the success of our STIS observation and data reduction strategy. Slight discrepancies may be due to differences in the instrumental resolving power, telluric correction residuals or uncertainties in the continuum level of the high-dispersion CFHT spectrum. Some uncertainty in the STIS scattered light correction also remains, leading to uncertainty in the zero level. More detailed characterization of the (2D) scattered light properties for STIS scan exposures awaits further study.

\begin{figure*}
\includegraphics[width=\textwidth]{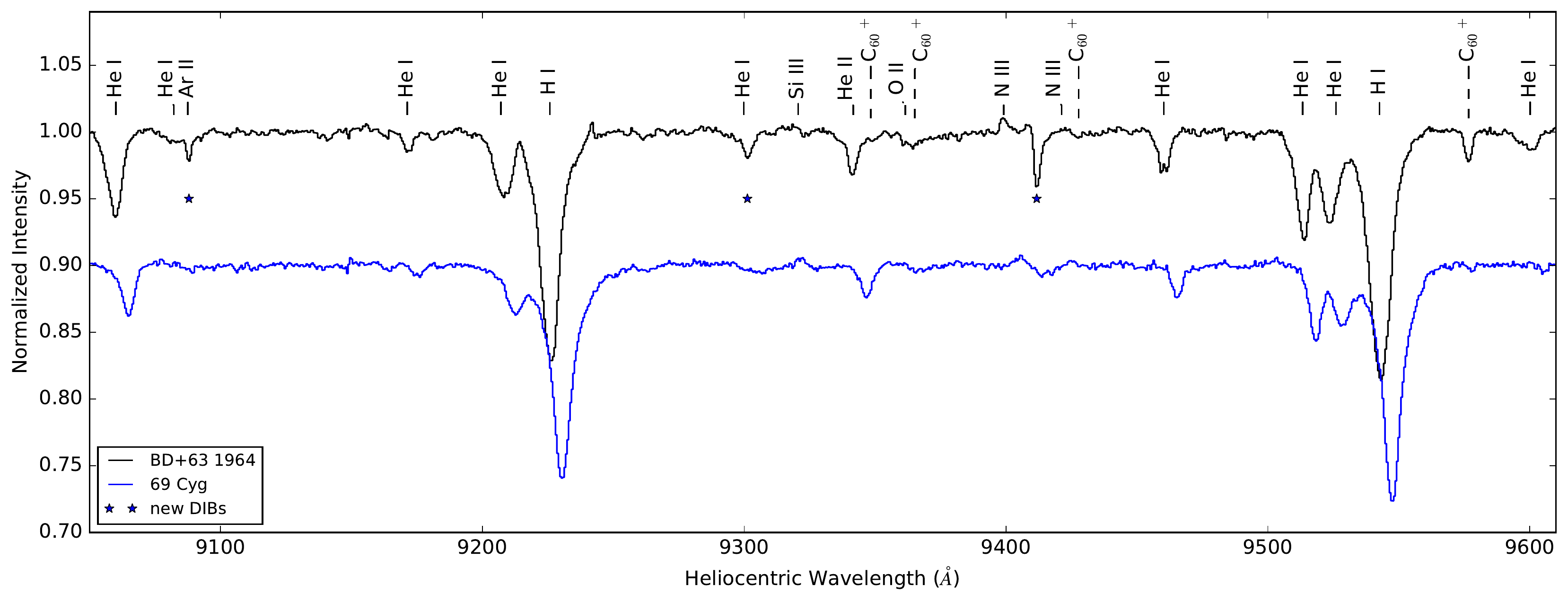}
\caption{STIS NIR spectra of \bd\ and 69~Cyg (offset vertically for display). Stellar features (in the rest frame of \bd) and the positions of the laboratory \cp\ bands (in the interstellar rest frame) are labeled. Possible new DIBs are marked with asterisks.\label{fig:spectra}}
\end{figure*}

\begin{figure*}
\includegraphics[width=\textwidth]{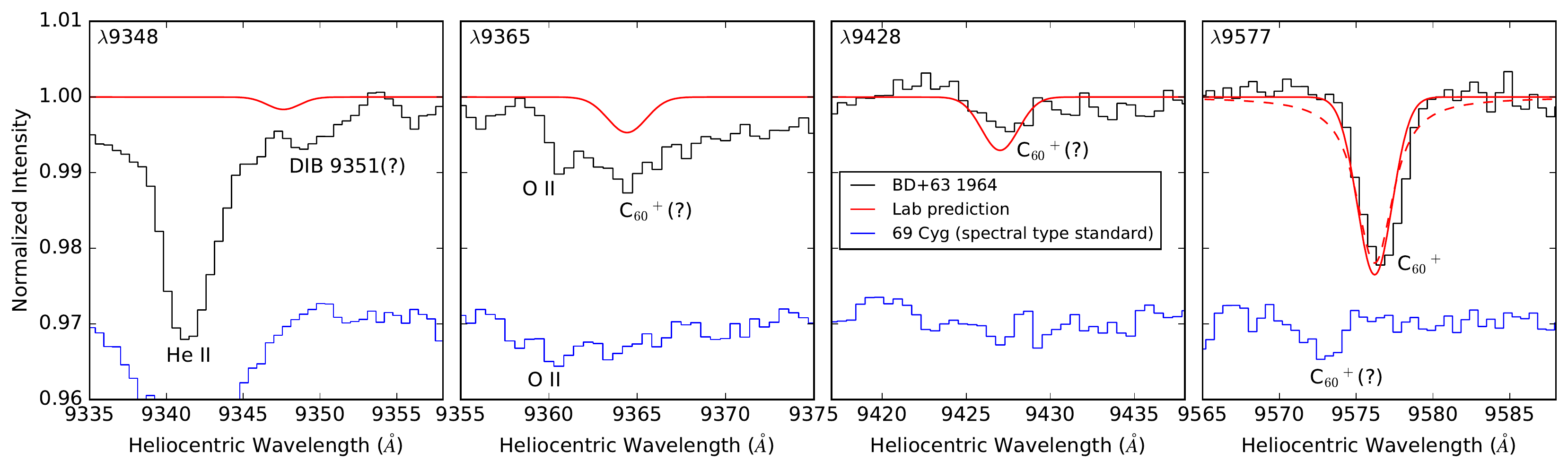}
\caption{STIS NIR spectra of \bd\ (black histograms) and the spectral-type standard 69 Cyg (blue histograms; shifted in wavelength to align the stellar features and offset vertically for display), for the four \cp\ bands in our spectral range. Red curves show predicted \cp\ DIB profiles based on \citet{cam16b}, Doppler-shifted to match the radial velocity of interstellar K\,{\sc i}  towards \bd. A Lorentzian model profile is also shown for \dib9577 (dashed line). { Detected and tentative (?) stellar and interstellar features are labeled.} \label{fig:zoom}}
\end{figure*}

\section{Results}\label{results}

The combined, scattered-light-subtracted exposures of \bd\ and 69~Cyg reached total counts of $8\times10^5$ and $5.5\times10^5$, respectively (per dispersion pixel). The presence of weak stellar and interstellar features precludes an accurate measurement of the continuum RMS, but in regions free of any obvious lines, S/N~=~600-800 was found for \bd\ and $\sim700$ for 69~Cyg. To our knowledge, this is the highest S/N ever demonstrated with STIS for direct near-infrared spectroscopy at full grating resolution. The continuum-normalized spectra are shown in Figure \ref{fig:spectra}.

Both stars are B0 supergiants and have nearly identical stellar spectra.  We used the non-LTE model atmosphere code CMFGEN \citep{hil98,hil99,hil11}, which solves the radiative-transfer equation for a spherically symmetric wind in the co-moving frame under the constraints of radiative and statistical equilibrium. { An effective temperature of 27,000~K, macro-turbulent velocity of 120~\kms\ and projected rotational velocity of 85~\kms\ was derived for both stars}.  The modeled lines are identified in Figure \ref{fig:spectra}. Four features stand out in the \bd\ spectrum that are not present in 69~Cyg or in the stellar model, which are likely due to interstellar absorption. The first three of these (at 9088, 9302 and 9412~\AA) are identified as new diffuse interstellar bands; \citet{gal00} also found a possible DIB near 9412~\AA. The absorption feature at 9577~\AA\ coincides with the laboratory \cp\ wavelength. { 69~Cyg also shows evidence for weak absorptions at 9577 and 9412~\AA}.

The equivalent width (EW) of the \dib9577 band is $74\pm3$~m\AA. A Gaussian fit to its profile gives EW~=~$76\pm3$~m\AA, with a central wavelength of $9576.60\pm0.03$~\AA, a central depth of 2.3\% and FWHM~=~$3.1\pm0.1$~\AA. These line strength measurements may be considered accurate to within about $\pm10$\% due to uncertainties in the STIS scattered light subtraction.

A close-up view of the four observed \cp\ band regions is shown in Figure \ref{fig:zoom}. A spectral model for \cp\ is overlaid using the laboratory wavelengths and band strengths of \citet{cam16}. We adopt Gaussian band shapes because a Lorentzian profile provides a poor match to the observed \dib9577 band wings (Fig. \ref{fig:zoom}). The model has been Doppler shifted to match the interstellar K\,{\sc i} velocity. As with most heavily-reddened diffuse sightlines, the K\,{\sc i} line towards \bd\ (recorded in our \espadons\ data) shows complex velocity structure, with three main components at $-16$, $-27$ and $-33$~\kms. We take the average of these ($-25$~\kms) as the interstellar \cp\ radial velocity. The \dib9577 band carrier abundance could be variable across the three K\,{\sc i} clouds (due to variations in chemical abundances and ionization levels), so we assign an error margin of $\pm10$~\kms, which corresponds to $\pm0.3$~\AA. Within the uncertainties, our inferred (Doppler corrected) \dib9577 rest wavelength of $9577.4$~\AA\ matches the laboratory value of $9577.0\pm0.2$~\AA; the slight redshift could indicate that \cp\ is more associated with the higher-velocity K\,{\sc i} gas. { The 1.0~\AA\ instrumental resolution contributes 0.2~\AA\ to the observed DIB FWHM.} The additional (0.4~\AA) of broadening in the observed band compared with the laboratory measurement of 2.5~\AA\ is consistent with the spread of interstellar K\,{\sc i} velocities in this sightline. 

We find weak absorption features in our \bd\ spectrum close to the wavelengths of the \dib9348, \dib9365 and \dib9428 \cp\ bands. These features are not present towards our standard star 69~Cyg, and may therefore be due to interstellar \cp. However, the appearance of the \dib9348 band is far from clear due to contamination { by another possible weak DIB at 9350~\AA\ (previously identified at a rest wavelenth of 9351~\AA\ by \citealt{wal16})}, as well as overlapping stellar He\,{\sc ii} absorption. The \dib9365 feature { is partially contaminated by a weak stellar O\,{\sc ii} line, and} is broader and deeper than expected. The \dib9428 feature is too weak for a reliable measurement.

\section{Discussion}

The strengths of different diffuse interstellar bands are known to vary among sightlines with differing physical and chemical properties, independent of the total amount of interstellar material \citep[\eg][]{kre87,ehr95,cam97,ens17,ely17}. The equivalent width per unit reddening (EW/\ebv) provides a measure of the relative strength of a given DIB. For \bd\, we find EW(9577)/\ebv~=~73~m\AA, which is among the smallest values known for this band. Unfortunately for our present study, this means that \bd\ is one of the least favorable sightlines in which to search for \cp, which explains to some degree our difficulties in establishing (or disproving) the presence of the weaker \cp\ absorption bands.

\begin{figure}
\includegraphics[width=\columnwidth]{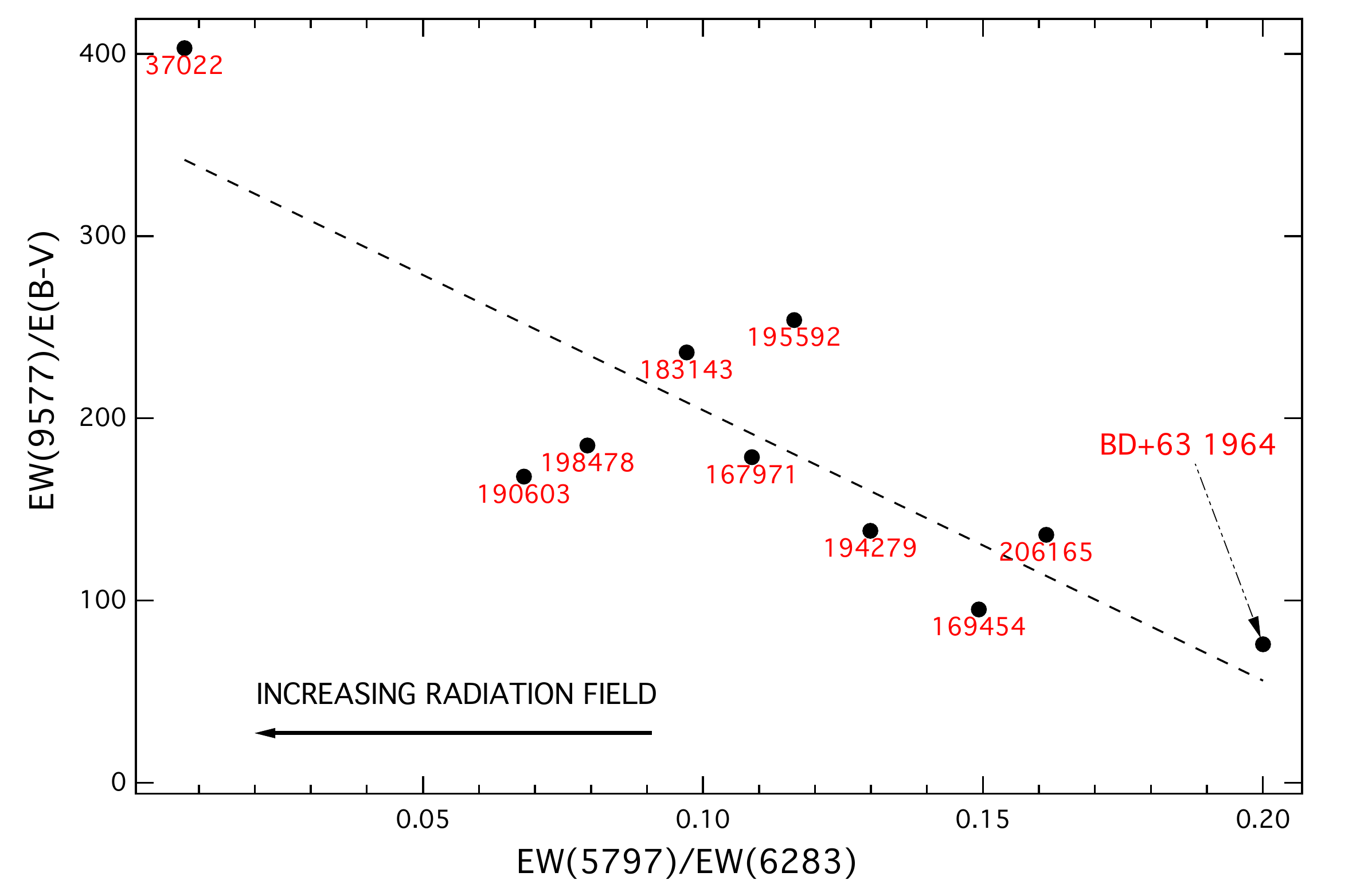}
\caption{Strength of the \dib9577 \cp\ band (per unit reddening) \emph{vs.} equivalent width ratio of the \dib6283 and \dib5797 DIBs (an indicator of radiation field strength), { for a sample of Galactic sightlines, labeled by HD number}. \label{fig:vs6283}}
\end{figure}

The extreme weakness of the \dib9577 band may be due to a lower than normal degree of ionization in this sightline.  The well-known DIB at 6283~\AA\ has a large EW/\ebv\ value in diffuse clouds that are strongly irradiated by UV (see above references), while the 5797~\AA\ DIB tends to be favored in more neutral, less irradiated clouds. The \dib5797/\dib6283 equivalent width ratio is thus considered to be a tracer of the radiation field strength in the line of sight. Data have been taken from compilations of optical DIB measurements by \citet{xia17} and the \dib9577 measurements of \citet{gal00} to study the relationship between EW(9577)/\ebv\ and the \dib5797/\dib6283 DIB strength ratio. Additional \dib6283 and \dib5797 measurements were obtained from archival data from the CFHT and Telescope Bernard Lyot. The results for 9 sightlines are shown in Figure~\ref{fig:vs6283}, including the present results for \bd. Even without HD\,37022 (a highly-irradiated sightline through the Orion Nebula with unusually high EW(9577)/\ebv), the trend is for increasing \dib9577 strength in gas pervaded by stronger radiation fields. In a survey of DIBs and atomic lines towards \bd\ by \citet{ehr97}, a relatively high gas density and low level of ionization was inferred, and the extremely low \dib5797/\dib6283 ratio is suggestive of weakly-irradiated gas. The trend for increasing EW(9577)/\ebv\ with radiation field strength is consistent with the assignment to \cp\ or another species whose ionisation stage becomes dominant in more irradiated diffuse clouds \citep[see also][]{foi97}.

\section{Conclusion}\label{conclusion}

We performed the first successful test of the HST `STIS scan' spectroscopic observational mode. The resulting spectroscopic S/N (up to $\sim800$) is, to our knowledge, by far the highest demonstrated with STIS for direct stellar measurements at full grating resolution. This technical demonstration opens up a new wavelength range for interstellar, stellar and exoplanetary spectroscopy, with unprecedented sensitivity.

We obtained the first `clean' measurements of the spectral region covering the \dib9348, \dib9365, \dib9428 and \dib9577 \cp\ bands, without the confounding presence of telluric contamination or fringing artifacts. The \dib9577 band exhibits a closely Gaussian shape with FWHM and peak wavelength consistent with laboratory measurements. However, due to the surprising weakness of the \cp\ features in our chosen sightline, the strengths and profiles of the (intrinsically weaker) \dib9348, \dib9365 or \dib9428 bands could not be reliably measured. We identify a correlation between the \dib9577 band strength (per unit reddening) and the \dib5797/\dib6283 DIB strength ratio (an indicator for radiation field strength), which suggests that the weakness of \dib9577 towards \bd\ is due to a low-radiation environment, consistent with the assignment of this DIB to an ion of a species whose first ionization energy is much less than that of neutral hydrogen (13.6~eV), and whose second ionization energy is $\gtrsim13.6$~eV. Using the STIS scanning technique, high-S/N HST spectroscopy of more strongly UV-irradiated, heavily reddened sightlines may provide the best opportunity for a conclusive identification of all five \cp\ bands, which will be required to place the detection of this molecule beyond reasonable doubt.

Possible new DIBs are reported at 9088, 9302 and 9412~\AA. Their interstellar origin may be confirmed by followup studies of stars with differing spectral types and degrees of extinction.

\acknowledgments
Based on observations made with the NASA/ESA Hubble Space Telescope (program \#14705). Support was provided by NASA through a grant from the Space Telescope Science Institute, which is operated by the Association of Universities for Research in Astronomy, Inc., under NASA contract NAS 5-26555. F.N. acknowledges Spanish grants FIS2012-39162-C06-01, ESP2013-47809-C3-1-R and ESP2015-65597-C4-1-R. \espadons\ observations were obtained under program 16BD86 at the Canada-France-Hawaii Telescope, which is operated by the National Research Council of Canada, the Institut National des Sciences de l'Univers of the Centre National de la Recherche Scientifique of France, and the University of Hawaii.



\end{document}